\documentclass[prb,twocolumn,showpacs,epsfig,epsf]{revtex4}
\usepackage{amsmath,amsthm,amsfonts,amscd,mathrsfs,pifont,bm}
\usepackage{eucal,graphicx}
\begin{document}

\title  {Random matrix theory of quantum transport in chaotic cavities with non-ideal leads}

\author {Andrzej Jarosz${}^{1}$, Pedro Vidal${}^{2}$ and Eugene Kanzieper${}^{3,4}$}

\affiliation
       {
       ${}^1$~Institut f\"ur Theoretische Physik, Universit\"at zu K\"oln, K\"oln 50937, Germany \\
       ${}^2$~Fakult\"at f\"ur Physik, Universit\"at Bielefeld, Bielefeld 33615, Germany\\
       ${}^3$~Department of Applied Mathematics, H.I.T.---Holon Institute of
       Technology, Holon 5810201, Israel\\
       ${}^4$~Department of Physics of Complex Systems, Weizmann Institute of
       Science, Rehovot 7610001, Israel
       }
\date   {\today}

\begin  {abstract}
We determine the joint probability density function (JPDF) of reflection eigenvalues in three Dyson's ensembles of normal-conducting chaotic cavities coupled to the outside world through both ballistic and tunnel point contacts. Expressing the JPDF in terms of hypergeometric functions of matrix arguments (labeled by the Dyson index $\beta$), we further show that reflection eigenvalues form a determinantal ensemble at $\beta=2$ and a new type of a Pfaffian ensemble at $\beta=4$. As an application, we derive a simple analytic expression for the concurrence distribution describing production of orbitally entangled electrons in chaotic cavities with tunnel point contacts when time reversal symmetry is preserved.
\end{abstract}

\maketitle
\newpage
{\it Introduction.}---Phase-coherent quantum transport \cite{I-2002} in irregular cavities with fully chaotic classical dynamics exhibits a remarkable statistical universality \cite{B-1997,A-2000}. Fluctuations of various transport observables (such as electrical or thermal conductance, noise power, transferred charge etc.) are described by universal statistical laws which appear to depend on the fundamental symmetries \cite{AZ-1997} of a cavity (such as the absence or presence of time-reversal, spin-rotational, particle-hole and chiral/sublattice symmetries) and the transmission properties of the leads attached to it. Other system-specific microscopic details of the scattering region become irrelevant after appropriate ensemble or energy averaging.

Out of three theoretical frameworks -- random matrices \cite{M-2004,F-2010}, semiclassical \cite{R-2000,W-2012} and field-theoretic approaches \cite{E-1997} -- devised for a non-perturbative description of transport phenomena in chaotic cavities, the random matrix theory is particularly well positioned to explore the universal aspects of quantum transport.

Since at low temperatures and voltages the transport properties can be related to the scattering matrix ${\bm {\mathcal S}}(\varepsilon_F)$ of the total system comprised by the cavity and the leads,
\begin{equation}
\label{hf}
    {\bm {\mathcal S}}(\varepsilon_F) = \openone_N - 2i \pi  {\bm {\mathcal W}}^\dagger
    ({\varepsilon}_F\,\openone_M - {\bm{\mathcal H}} + i \pi {\bm{\mathcal W}}
    {\bm {\mathcal W}}^\dagger)^{-1} {\bm{\mathcal W}},
\end{equation}
it becomes a central object of interest. Equation (\ref{hf}), known as the Heidelberg formula \cite{MW-1969}, suggests that the symmetries and statistics of the scattering matrix derive from the symmetries and statistics of an $M \times M$ random-matrix-Hamiltonian ${\bm {\mathcal H}}$ used to mimic a chaotic scattering of a single electron inside the cavity as $M\rightarrow \infty$. The coupling of electron states with the Fermi energy ${\varepsilon_F}$ in the cavity to those in the leads is described by an $M\times N$ deterministic matrix ${\bm {\mathcal W}}$, where $N=n_{\rm L} + n_{\rm R}$ is the total number of
propagating modes (channels) in the left ($n_{\rm L}$) and right ($n_{\rm R}$) leads.

In {\it normal}-conducting cavities with broken time-reversal symmetry the $N\times N$ scattering matrix ${\bm {\mathcal S}}={\bm {\mathcal S}}(\varepsilon_F)$ is merely unitary ($\beta=2$); it becomes unitary symmetric ${\bm {\mathcal S}}={\bm {\mathcal S}}^{\rm T}$ when both time-reversal and spin-rotational symmetries are preserved ($\beta=1$), and unitary self-dual quaternion ${\bm {\mathcal S}}= {\bm \sigma}_y {\bm {\mathcal S}}^{\rm T}{\bm \sigma}_y$ if the time-reversal symmetry is preserved but the spin-rotational one is broken ($\beta=4$). Fluctuations of the scattering matrix are described by the Poisson kernel \cite{H-1963,B-1995,MB-1999}
\begin{equation}\label{pk}
    \Pi^{(\beta)}_{\bm{{\mathcal S}_0}}(\bm{\mathcal S}) \propto
    \big[\det{} (\openone_N - \bm{{\mathcal S}_0}^\dagger \bm{{\mathcal S}_0})\big]^{\theta_N^{(\alpha)}}
    \big| \det{}(\openone_N-\bm{{\mathcal S}_0}^\dagger \bm{\mathcal S})\big|^{-2\theta_N^{(\alpha)}}.
\end{equation}
Here $\alpha=2/\beta$ is a complementary symmetry parameter, and $\theta_N^{(\alpha)} = (N-1)/\alpha+1$. The notation ``${\rm det}$'' stands for a conventional determinant for $\beta=1$ and $2$; it should be interpreted as a quaternion determinant \cite{M-2004} for $\beta=4$. Equation (\ref{pk}) highlights the universal character of scattering matrix fluctuations: it depends on the fundamental system symmetries encoded in the Dyson index $\beta$ and parameters of point contacts contained in the average scattering matrix $\bm{{\mathcal S}_0}$. The $(n_{\rm L}+n_{\rm R})$ eigenvalues $\hat{\bm \gamma} = (\hat{\bm \gamma}_{\rm L}, \hat{\bm \gamma}_{\rm R})={\rm diag}\left(\{ \sqrt{1-\Gamma_j}\}\right)$ of ${\bm {\mathcal S}_0}$ characterise\cite{B-1997} couplings between the cavity and the leads in terms of tunnel probabilities $\Gamma_j$ of the $j$-th electron mode in the leads.

In chaotic cavities probed via {\it ballistic} point contacts \cite{vHB-1996} (so-called `{\it ideal} leads' characterized by $\Gamma_j=1$), the average scattering matrix vanishes and the Poisson kernel degenerates to the uniform distribution. The latter implies that fluctuations of scattering matrices are described by three Dyson's circular ensembles \cite{M-2004} -- COE ($\beta=1$), CUE ($\beta=2$) or CSE ($\beta=4$). The uniformity, in turn, induces nontrivial correlations \cite{BM-1994} between reflection eigenvalues ${\bm R}=(R_1,\dots,R_n)$ which, together with appropriate unitary rotations [see Eq.~(\ref{dU}) below], conveniently parameterize the scattering matrix:
\begin{equation}
\label{PnT}
    P_{({\bm 0}|{\bm 0})}^{(\beta)}({\bm R}) \propto  \,|\Delta_n^\beta({\bm R})| \prod_{j=1}^n
    (1-R_j)^{\beta/2-1 + \beta \nu /2}.
\end{equation}
This JPDF refers to chaotic cavities probed via two ideal leads. Here, $\Delta_n({\bm R})=\prod_{j<k} (R_k-R_j)$ is the Vandermonde
determinant with $n = {\rm min}(n_{\rm L},n_{\rm R})$, and the parameter $\nu= |n_{\rm L}-n_{\rm R}|$ accounts for asymmetry between the leads.

Considered though the prism of Landauer-type formulae \cite{LFLB-1957}, Eq.~(\ref{PnT}) appears to be of invaluable operational importance. Based on it, detailed non-perturbative predictions \cite{B-1997,B-2011} have been made for statistics of electrical conductance \cite{BM-1994,B-1997,SSW-2008,KSS-2009}, noise power \cite{B-1997,SS-2006,SSW-2008,KSS-2009} and transferred charge \cite{SSW-2008,N-2008} and a surprising link between zero-dimensional theories of quantum transport and the theory of integrable lattices \cite{OK-2008,MS-2013,K-2014} has been unveiled.

Recently, the interest in random matrix theories of quantum transport has been revived. The advent of topological superconductors \cite{TSC-2008} prompted a series of works \cite{DBB-2010,M-2002} where novel {\it circular} ensembles \cite{B-2014} of scattering matrices (complementary to Dyson's COE, CUE and CSE) were defined to reveal universal features of transport phenomena in irregular structures described by random-matrix Hamiltonians of the Bogoliubov-de Gennes and Dirac types. Among other results, various extensions of Eq.~(\ref{PnT}) were derived therein \cite{DBB-2010,M-2002,B-2014} in the context of thermal and electrical conductances.

While much progress has been achieved in a non-perurbative description of chaotic cavities probed via ballistic point contacts, the world of {\it non-ideal (tunnel) couplings} is barely understood from the random-matrix-theory perspective. (Involved mathematical structures \cite{H-1963,VK-2012} lurking behind the Poisson kernel \cite{B-1995,B-2009} are at the root of the poor knowledge.) This is in stark contrast to remarkable technological developments in the field: chaotic structures with adjustable point contacts were fabricated \cite{KMR-1997} long ago, and a possibility to efficiently control degree of dephasing \cite{TD-papers} was recently demonstrated \cite{RPRCFGM-2009}.

{\it Effect of tunnel point contacts on JPDF of reflection eigenvalues.}---In this paper, we investigate the influence of non-ideal couplings on statistics of reflection eigenvalues in normal-conducting chaotic cavities belonging to either of the three Dyson symmetry classes ($\beta=1,\,2$ or $4$). Focussing on the two-lead geometry and assuming that only one (left) lead is attached via tunnel point contacts, we shall show that the corresponding JPDF is given by the expression
\vspace{-1cm}
\begin{widetext}
\begin{equation}\label{jpdf-1}
   P^{(\beta)}_{(\hat{{\bm \gamma}}_{\rm L}|\,{\bm 0})}({\bm R}) \propto
   \big[{\rm det}{} (\openone_{n_{\rm L}} - \hat{{\bm \gamma}}_{\rm L}^2) \big]^{\theta_N^{(\alpha)}}
    \, \big|\Delta_{n}^{\beta} ({\bm R})\big| \,
    \prod_{j=1}^{n} (1-R_j)^{\beta\nu/2+\beta/2-1}
    \;\cdot
    {}_2 {\mathcal F}_1^{(\alpha)}
    \left(\begin{array}{c}
              \theta_{N}^{(\alpha)},\, \theta_{N}^{(\alpha)} \\
              \theta_{n_{\rm L}}^{(\alpha)}
            \end{array}\right| \left. \begin{array}{c}
              {} \\
              {}
            \end{array} \hat{\bm{\gamma}}_{\rm L}^2,\, \bm{R}_*\right).
\end{equation}
\end{widetext}
Here, ${}_2 {\mathcal F}_1^{(\alpha)}(\;\cdot\;| {\bm X},{\bm Y})$ is a hypergeometric function \cite{M-1995,F-2010} of two matrix arguments, ${\bm X} = {\rm diag}(x_1,\dots,x_M)$ and ${\bm Y} = {\rm diag}(y_1,\dots,y_M)$; it is defined by the series
\begin{eqnarray}\label{hf2ma}
    {}_2 {\mathcal F}_1^{(\alpha)}
\left(\begin{array}{c}
              a_1, a_2 \\
              b_1
            \end{array}\right| \left. \begin{array}{c}
              {} \\
              {}
            \end{array} {\bm X},\, {\bm Y}\right) = \qquad\qquad\qquad\qquad\qquad\nonumber\\
            \sum_{\{\lambda:\,\ell(\lambda)\le M\}} \frac{1}{|\lambda|!}
            \frac{[a_1]_\lambda^{(\alpha)}  [a_2]_\lambda^{(\alpha)}}{[b_1]_\lambda^{(\alpha)}}
            \frac{C_\lambda^{(\alpha)} ({\bm X})\, C_\lambda^{(\alpha)} ({\bm Y})}{C_\lambda^{(\alpha)} (\openone_n)}
\end{eqnarray}
running over ordered, non-increasing partitions $\lambda = (\lambda_1,\lambda_2,\dots)$ whose length $\ell(\lambda)$ does not exceed $M$; $|\lambda|$ is weight of $\lambda$; a generalized Pochhammer symbol $[a]_\lambda^{(\alpha)}$ equals
\begin{eqnarray}
    [a]_\lambda^{(\alpha)} = \prod_{j=1}^{\ell(\lambda)} \left(
        a - \frac{j-1}{\alpha}
    \right)_{\lambda_j},\quad (a)_k = \frac{\Gamma(a+k)}{\Gamma(a)}.
\end{eqnarray}
The function $C_\lambda^{(\alpha)} ({\bm X})$ is the Jack polynomial \cite{M-1995} in the $C$-normalization \cite{F-2010}, such that $\sum_{|\lambda|=k} C_\lambda^{(\alpha)} ({\bm X}) = ({\rm tr}{\bm X})^k$.
The vector $\hat{\bm{\gamma}}_{\rm L}^2=(1-\Gamma_1,\dots, 1-\Gamma_{n_{\rm L}})$ accommodates a set of tunnel probabilities $(\Gamma_1,\dots,\Gamma_{n_{\rm L}})$ quantifying a non-ideal coupling of the left lead; the vector ${\bm R}_* ={\bm R}$ for $n_{\rm L} \le n_{\rm R}$, and ${\bm R}_* = ({\bm R} \cup \openone_{n_{\rm L}-n_{\rm R}})$ otherwise. Equation (\ref{jpdf-1}) is our {\it first main result}.

For generic $\beta$, Eq.~(\ref{jpdf-1}) can be written in a more informative form provided that tunnel probabilities are channel independent, $\hat{\bm{\gamma}}_{\rm L}^2 = \gamma^2 \openone_{n_{\rm L}}$. In this case, a hypergeometric function of two matrix arguments reduces to that of {\it one} matrix argument,  ${}_2 {\mathcal F}_1^{(\alpha)}(\,\cdot\,|\,{\bm X},\openone_M)={}_2 F_1^{(\alpha)}(\,\cdot\,| \,{\bm X})$. The latter is related to the Selberg correlation integral \cite{K-1993}:
\begin{eqnarray}\label{sci}
    S_{n,m}^{(\sigma)}(\lambda_1,\lambda_2; {\bm x}) = \int_0^1 dt_1 \cdots \int_0^1 dt_n\,\qquad\qquad\qquad \nonumber \\
    \times |\Delta_n^\sigma({\bm t})|
        \prod_{j=1}^n
        \left( t_j^{\lambda_1} (1-t_j)^{\lambda_2}
        \prod_{k=1}^m (t_j-x_k)\right),
\end{eqnarray}
where ${\bm x} = (x_1,\dots,x_m)$. Straightforward calculations yield \cite{JVK-2014}:
\begin{eqnarray}\label{jpdf-2}
   P^{(\beta)}_{(\hat{{\bm \gamma}}_{\rm L}|\,{\bm 0})}({\bm R}) \propto
   (1-\gamma^2)^{n_{\rm L} \theta_N^{(\alpha)} }
   \left[{\rm det}{} \left(\frac{\openone_{n_{\rm L}}}{\openone_{n_{\rm L}} - \gamma^2 {\bm R}_*}\right) \right]^{\theta_N^{(\alpha)}}
   \nonumber\\
   \times S_{n_{\rm R}/\alpha,n_{\rm L}}^{(4/\beta)}
    \left(
    \frac{2}{\beta}-1, \frac{2}{\beta}-1; \frac{\openone_{n_{\rm L}}}{\openone_{n_{\rm L}} - \gamma^2 {\bm R}_*}
    \right) \, P^{(\beta)}_{({\bm 0}|\,{\bm 0})}({\bm R}).\nonumber\\{}
\end{eqnarray}
Here, $n_{\rm R}/\alpha$ is assumed to be an integer; for $\beta=1$ symmetry class, this restricts $n_{\rm R}$ to a set of even integers.

Remarkably, Eq.~(\ref{jpdf-2}) allows us to derive a {\it pfaffian} representation of the JPDF of reflection eigenvalues at $\beta=4$, which is complementary to a previously established \cite{VK-2012} {\it determinantal} representation of the JPDF at $\beta=2$. Indeed, at $\beta=4$, the Selberg correlation integral $S_{2n_{\rm R},n_{\rm L}}^{(1)}$ is essentially an average product of characteristic polynomials in the Jacobi {\it orthogonal} ensemble. Applying methods detailed in Ref. \cite{BS-2006}, we derive \cite{JVK-2014}:
\begin{eqnarray}\label{b4-1}
    P^{(4)}_{(\gamma \openone_{n_{\rm L}}|\,{\bm 0})}({\bm R}) &\propto&
     \prod_{j=1}^{n} (1-R_j)^{2\nu + 1} \,\nonumber\\
     &\times& \Delta_{n}^{3} ({\bm R}) \,
     {\rm pf} \left[ W_\gamma^{(4)}(R_j,R_k)\right],
\end{eqnarray}
where
\begin{eqnarray}\label{b4-2}
    W_\gamma^{(4)}(R_j,R_k) &=& (R_j-R_k)\nonumber\\
        &\times& {}_2 F_1^{(1/2)}
    \Bigg(\begin{array}{c}
              a, a  \\
              3
            \end{array}\Big|  \begin{array}{c}
              {} \\
              {}
            \end{array} \gamma^2 \left(\begin{array}{cc}
                                   R_j & 0 \\
                                   0 & R_k
                                 \end{array}\right)
            \Bigg)\qquad
\end{eqnarray}
is an antisymmetric two-point scalar kernel expressed in terms of a hypergeometric function of a $2\times 2$ matrix argument; $a=2n_{\rm R}+n_{\rm L}+1$. Alternatively, the kernel $W_\gamma^{(4)}(R_j,R_k)$ can be written \cite{JVK-2014} in terms of Jacobi polynomials. This result holds true for $n_{\rm L}$ even and $n_{\rm L} \le n_{\rm R}$; a generic case will be reported elsewhere. Equations (\ref{b4-1}) and (\ref{b4-2}) represent our {\it second main result}. To the best of our knowledge, the Pfaffian ensemble ``$\Delta^3 \, {\rm pf}(\cdot)$'' has never been reported in the literature. Finding its correlation functions appears to be a nontrivial problem.

Finally, let us comment on the $\beta=1$ symmetry class, where one could na\"ively anticipate appearance of an average product of characteristic
polynomials in the Jacobi {\it symplectic} ensemble. To realize that this is {\it not} the case, we turn to Eq.~(\ref{jpdf-2}) to observe that the JPDF is now given by the Selberg correlation integral $S_{n_{\rm R}/2,n_{\rm L}}^{(4)}$ with $n_{\rm R}$ kept even. Consulting Eq. (\ref{sci}), we readily conclude that it defines an average of a {\it square root} of the product of characteristic polynomials. Algebraic structures behind these objects are scarcely studied \cite{FN-2014}.

{\it Concurrence distribution at $\beta=1$.}---The JPDF in the form Eq.~(\ref{jpdf-1}) lays a basis for a nonperturbative analysis of various quantum transport effects \cite{JVK-2014}. As an illustration, let us turn to the problem of orbital entanglement production in a chaotic cavity with preserved time-reversal symmetry.

The proposal to use a cavity as an effective orbital entangler for pairs of non-interacting electrons was put forward by Beenakker and collaborators \cite{B-2004} a decade ago. Following these authors, we consider a cavity connected to an electron reservoir at the left and right through two pairs of single channel point contacts. Chaotic scattering entangles the outgoing state in the left channels with that in the right channels. The degree of entanglement is quantified by the concurrence ${\mathcal C} = 2 \sqrt{{\rm det} ({\bm \pi}{\bm \pi}^\dagger)}/{{\rm tr}({\bm \pi}{\bm \pi}^\dagger)}$, where ${\bm \pi}$ is the $2\times 2$ matrix ${\bm \pi} = {\bm \sigma}_y {\bm r} {\bm \sigma}_y {\bm t}^{\rm T}$. The concurrence nullifies in absence of entanglement and reaches unity for maximally entangled states.

Since single-channel point contacts drive the device into extreme quantum limit, fluctuations in entanglement production are expected to be very strong. Indeed, in the case of {\it ballistic} point contacts, it was found \cite{B-2004} that the average concurrence $\langle {\mathcal C}\rangle = \log (4/e)\approx 0.3863$ compares with its standard deviation $\langle\!\langle {\mathcal C} \rangle\!\rangle=\sqrt{2-(\log 4)^2} \approx 0.2796$. The higher-order cumulants required for an adequate description of concurrence fluctuations can be extracted from the probability density \linebreak $f_0({\mathcal C}) = 2/(1+{\mathcal C})^2$ calculated in Ref. \cite{GF-2008}.

The effect of {\it tunnel} point contacts on concurrence fluctuations has been studied numerically in Ref. \cite{AS-2010} where it was argued
that the orbital entanglement production can be optimised by increasing asymmetry between transparencies of left and right point contacts. Simulations of concurrence distribution for an entangler with a pair of ballistic and a pair of tunnel point contacts indicated that
a family of probability densities $\{f_\gamma({\mathcal C})\}$ is likely to exhibit a point of intersection at ${\mathcal C}={\mathcal C}_*\approx 1/3$; compared to the case of ideal couplings, the probability density was enhanced for ${\mathcal C}>{\mathcal C}_*$ and diminished for ${\mathcal C}<{\mathcal C}_*$.

The JPDF of reflection eigenvalues Eq.~(\ref{jpdf-1}) makes it possible \cite{VV-2013} to support this observation analytically. Expressing the concurrence in terms of reflection eigenvalues,
\begin{eqnarray}
{\mathcal C} = 2\frac{\sqrt{R_1(1-R_1)R_2(1-R_2)}}{R_1+R_2-2 R_1R_2},
\end{eqnarray}
and deducing from Eq.~(\ref{jpdf-2}) that
\begin{eqnarray}
    P^{(1)}_{(\gamma \openone_2|\,{\bm 0})}(R_1,R_2) = \frac{8}{3\pi} (1-\gamma^2)^5
    |R_1-R_2|
    \qquad \nonumber\\
    \qquad\times \frac{1+ \frac{2}{3}\gamma^2 (R_1+R_2) +\gamma^4 R_1 R_2}{\prod_{j=1}^2  R_j^{1/2}(1-R_j)^{1/2}(1-\gamma^2 R_j)^{7/2}},
\end{eqnarray}
we derive after some algebra:
\begin{eqnarray}\label{fgamma}
    f_\gamma({\mathcal C}) = f_0({\mathcal C}) + \frac{2\gamma^4}{(1+{\mathcal C})^3}\left(
    {\mathcal C} - \frac{1}{3}
    \right).
\end{eqnarray}
This result locates the intersection point at ${\mathcal C}_* = 1/3$ corroborating numerics of Ref. \cite{AS-2010} extremely well.

{\it Sketch of the derivation (JPDF).}---The probability density of $n_{\rm L}\times n_{\rm L}$ reflection matrix ${\bm r}$ is the starting point of our analysis. Substituting the average scattering matrix of the form ${{\bm {\mathcal S}}}_0 = (\hat{{\bm \gamma}}_{\rm L} \cup 0 \times \openone_{n_{\rm R}})$ into Eq.~(\ref{pk}) and eliminating `transmission' degrees of freedom, one obtains:
\begin{equation}
   {\tilde P}^{(\beta)}_{(\hat{{\bm \gamma}}_{\rm L}|\,{\bm 0})}({\bm r}) \propto
   \big[\det{} (\openone_{n_{\rm L}} - \hat{{\bm \gamma}}_{\rm L}^2)\big]^{\theta_N^{(\alpha)}}
    \big| \det{}(\openone_{n_{\rm L}}-\hat{{\bm \gamma}}_{\rm L} \bm{r})\big|^{-2\theta_N^{(\alpha)}}\nonumber
\end{equation} \vspace{-0.5cm}
\begin{equation}
    \times
    \big[ \det{}(\openone_{n_{\rm L}}-\bm{r}\bm{r}^\dagger)\big]^{\beta/2-1+\beta\nu/2}
    \theta(\openone_{n_{\rm L}} - {\bm r}{\bm r}^\dagger).
\end{equation}
Reflection eigenvalues 
are brought into play through a singular value decomposition ${\bm r} = {\bm u}  \hat{\bm \varrho} \,{\bm v}$, where $\hat{\bm \varrho} = {\bm R}^{1/2}_*$, whilst ${\bm u}$ and ${\bm v}$ are either constrained ($\beta=1,\,4$) or unconstrained ($\beta=2$) unitary matrices as specified by Eq.~(\ref{dU}) below. To determine the JPDF of reflection eigenvalues, one has to integrate out rotational degrees of freedom with respect to the weighted product of two Haar measures on $U(n_{\rm L})$,
\begin{eqnarray}
\label{dU}
    d\mu^{(\beta)}({\bm u},{\bm v}) =
     d\mu({\bm u})\, d\mu({\bm v}) \,
     \Bigg\{
    \begin{array}{cl}
    \delta({\bm v}-{\bm u}^{\rm T}), & \beta=1; \\
    1, & \beta=2;\\
    \delta({\bm v}-{\bm \sigma}_y {\bm u}^{\rm T} {\bm \sigma}_y),  & \beta=4. \quad
\end{array}
\end{eqnarray}
This yields:
\begin{eqnarray}\label{jpdf-int}
   P^{(\beta)}_{(\hat{{\bm \gamma}}_{\rm L}|\,{\bm 0})}({\bm R}) &=&
   \big[{\rm det}{} (\openone_{n_{\rm L}} - \hat{{\bm \gamma}}_{\rm L}^2) \big]^{\theta_N^{(\alpha)}}
    \, P^{(\beta)}_{({\bm 0},{\bm 0})}({\bm R})  \nonumber\\
    &\times&
    \int \frac{d\mu^{(\beta)}({\bm u},{\bm v})}{\big| \det{}(\openone_{n_{\rm L}}-\hat{{\bm \gamma}}_{\rm L} {\bm u} \hat{\bm{\varrho}}\, {\bm v})\big|^{2\theta_N^{(\alpha)}}}.
\end{eqnarray}
The group integral \cite{G-int},
\begin{equation}\label{G-i}
    I^{(\beta)}_{(p,q)} ({\bm A},{\bm B}) = \int \frac{d\mu^{(\beta)}({\bm u},{\bm v})}{\big| \det{}(\openone_{p}-{\bm A} {\bm u} {\bm B}\, {\bm v})
    \big|^{2q}},
\end{equation}
appearing in Eq.~(\ref{jpdf-int}), has previously been calculated \cite{O-2004,VK-2012} for $\beta=2$ only. To evaluate it for other $\beta$'s, we expand an inverse determinant
\begin{equation}\label{det}
    \det{}^{-q}(\openone_{p}-{\bm A} {\bm u} {\bm B} {\bm v}) =
    \sum_{\{\lambda,\,\ell(\lambda)\le p\}} \frac{[q]_\lambda^{(\alpha)}}{|\lambda|!} \, C_\lambda^{(\alpha)} ({\bm A} {\bm u} {\bm B} {\bm v})
\end{equation}
in terms of Jack polynomials $C_\lambda^{(\alpha)}$ to arrive at the series
\begin{equation}
    I^{(\beta)}_{(p,q)} ({\bm A},{\bm B})=
    \sum_{
    \begin{array}{c}
      {}_{\{\lambda,\,\ell(\lambda)\le p\}} \\
      {}_{\{\mu,\,\ell(\mu)\le p\}}
      \end{array}
    }
    \frac{[q]_\lambda^{(\alpha)} [q]_\mu^{(\alpha)} }{|\lambda|!\, |\mu|!}
    \, {\tilde I}^{(\beta)}_{\lambda,\mu} ({\bm A}, {\bm A}; {\bm B}),
\end{equation}
where
\begin{eqnarray}
    {\tilde I}^{(\beta)}_{\lambda,\mu} ({\bm A}_1,{\bm A}_2; {\bm B})\qquad \qquad\qquad\qquad \qquad\qquad\quad \nonumber\\
    =
    \int d\mu^{(\beta)}({\bm u},{\bm v})\,
    C_\lambda^{(\alpha)} ({\bm A}_1 {\bm u} {\bm B} {\bm v})\, \overline{C_\mu^{(\alpha)} ({\bm A}_2 {\bm u} {\bm B} {\bm v})}.\,
\end{eqnarray}
The latter integral can be calculated using insights from Ref. \cite{FS-2009}. Deferring details of the proof to a separate publication \cite{JVK-2014}, we state the result:
\begin{eqnarray}\label{fs-extended}
    {\tilde I}^{(\beta)}_{\lambda,\mu}
    =  \delta_{\lambda\mu}
    \frac{|\lambda|!}{[\theta_p^{(\alpha)}]_\lambda^{(\alpha)}}
    \frac{C_\lambda^{(\alpha)} ({\bm B} {\bm B}^\dagger)}{C_\lambda^{(\alpha)} (\openone_p)}\, C_\lambda^{(\alpha)} ({\bm A}_1 {\bm A}_2^\dagger).
\end{eqnarray}
Combining Eqs.~(\ref{hf2ma}), (\ref{G-i}) -- (\ref{fs-extended}) one concludes that
\begin{eqnarray}
    I^{(\beta)}_{(p,q)} ({\bm A},{\bm B})=
    {}_2 {\mathcal F}_1^{\,(\alpha)}
\left(\begin{array}{c}
              q,\, q \\
              \theta_p^{(\alpha)}
            \end{array}\right| \left. \begin{array}{c}
              {} \\
              {}
            \end{array} {\bm A}{\bm A}^\dagger,\, {\bm B}{\bm B}^\dagger\right).
\end{eqnarray}
Substituting it back to Eq.~(\ref{jpdf-int}) completes our derivation of the JPDF of reflection eigenvalues [Eq.~(\ref{jpdf-1})].

{\it Summary.}---In this paper, we formulated a random-matrix-theory approach to quantum transport in normal-conducting chaotic structures probed through both ballistic and tunnel point contacts. Starting with the Poisson kernel, we calculated the JPDF of reflection eigenvalues for arbitrary Dyson's index $\beta$ and showed that it is expressed in terms of a hypergeometric function of matrix arguments. This general result implies that reflection eigenvalues form a ``$\Delta\, {\rm det}(\cdot)$'' determinantal process for $\beta=2$ and a novel ``$\Delta^3\, {\rm pf}(\cdot)$'' pfaffian process for $\beta=4$. Although finding an algebraically insightful representation for the JPDF at $\beta=1$ remains an open problem, a hypergeometric representation of JPDF still is of operational value: as an example, we calculated the concurrence distribution for orbital entanglement production in chaotic cavities with asymmetric left and right point contacts.

Finally, let us point out that a general framework outlined in the paper lays a basis for many more -- both mathematics- and physics--oriented -- studies. In particular, analysis \cite{V-2014} of quantum transport effects in superconducting chaotic cavities \cite{DBB-2010,B-2009} and an exact solution \cite{JVK-2014} of B\"uttiker's dephasing model \cite{TD-papers} will be reported elsewhere.

{\it Acknowledgments.}---We thank F. A. G. Almeida for providing us with numerical data reported in Ref. \cite{AS-2010}. This research was supported by the Israel Science Foundation through the grant No 647/12 and by the German Research Foundation (DFG) through IRTG 1132.


\begin{references}
\bibitem{G-int} In Eq.~(\ref{G-i}) and below, matrices ${\bm A}$ and ${\bm B}$ are assumed to be arbitrary complex ($\beta=2$), complex symmetric ($\beta=1$), or self-dual quaternion ($\beta=4$).
\end{references}
\end{document}